\documentclass[10pt,journal]{IEEEtranTCOM}

\usepackage[english]{babel}
\usepackage[usenames]{color}
\usepackage[cp1250]{inputenc}
\usepackage{amsfonts}
\usepackage{amsthm}
\usepackage{graphicx}
\usepackage{epsfig}
\usepackage{mathrsfs}
\usepackage{amsmath}
\usepackage{algorithm}
\usepackage{algorithmic}
\usepackage{hyperref}

\theoremstyle{plain}

\begin{document}
\newcommand{\bea}{\begin{eqnarray}}
\newcommand{\eea}{\end{eqnarray}}
\newcommand{\be}{\begin{equation}}
\newcommand{\ee}{\end{equation}}
\newcommand{\beas}{\begin{eqnarray*}}
\newcommand{\eeas}{\end{eqnarray*}}
\newcommand{\bs}{\backslash}
\newcommand{\bc}{\begin{center}}
\newcommand{\ec}{\end{center}}
\def\SC {\mathscr{C}}
\newcommand{\tb}{\tilde{t}}
\newcommand{\xb}{\tilde{x}}
\newcommand{\vb}{\tilde{v}}

\title{Phase space maximal entropy random walk:\\ Langevin-like ensembles of physical trajectories}
\author{\IEEEauthorblockN{Jarek Duda}\\
\IEEEauthorblockA{Jagiellonian University,
Golebia 24, 31-007 Krakow, Poland,
Email: \emph{dudajar@gmail.com}}}
\maketitle

\begin{abstract}
As written by statistician George Box "All models are wrong, but some are useful", standard diffusion derivation or Feynman path ensembles use nonphysical infinite velocity/kinetic energy nowhere differentiable trajectories - what seems wrong, might be only our approximation to simplify mathematics. This article introduces some basic tools to investigate this issue. To consider ensembles of more physical finite velocity trajectories, we can work in $(x,v)$ phase space like in Langevin equation with velocity controlling spatial steps, here also controlled with spatial potential $V(x)$. There are derived and compared 4 approaches to predict stationary probability distributions: using Boltzmann ensemble of steps/points in space (GRW - generic random walk) or in phase space (psGRW), and analogously Boltzmann ensemble of paths in space (MERW - maximal entropy random walk) and in phase space (psMERW), also generalized to L{\'e}vy flights. Path ensembles generally have much stronger Anderson-like localization, MERW has stationary distribution exactly as quantum ground state. Proposed novel MERW in phase space has some slight differences, which might be distinguished experimentally. For example for 1D infinite potential well: $\rho=1$ stationary distribution for step ensemble, $\rho\sim \sin^2$ for path ensemble (as in QM), and $\rho\sim  \sin$ for proposed smooth path ensembles - more frequently approaching the barriers due to randomly gained velocity.  
\end{abstract}
\section{Introduction}
In standard derivation of diffusion equation, or in Feynman path integrals~\cite{feynman}, we consider steps in space, in continuous limit using $\epsilon\propto \delta^2\to 0$ for $\epsilon$ being temporal step and $\delta$ being spatial step - because width of Gaussian grows with square root of the number of steps. It means velocity and $mv^2/2$ kinetic energy goes to infinity ($v=\delta/\epsilon \to \infty$), nowhere differentiable trajectories. 

\begin{figure}[t!]
    \centering
        \includegraphics{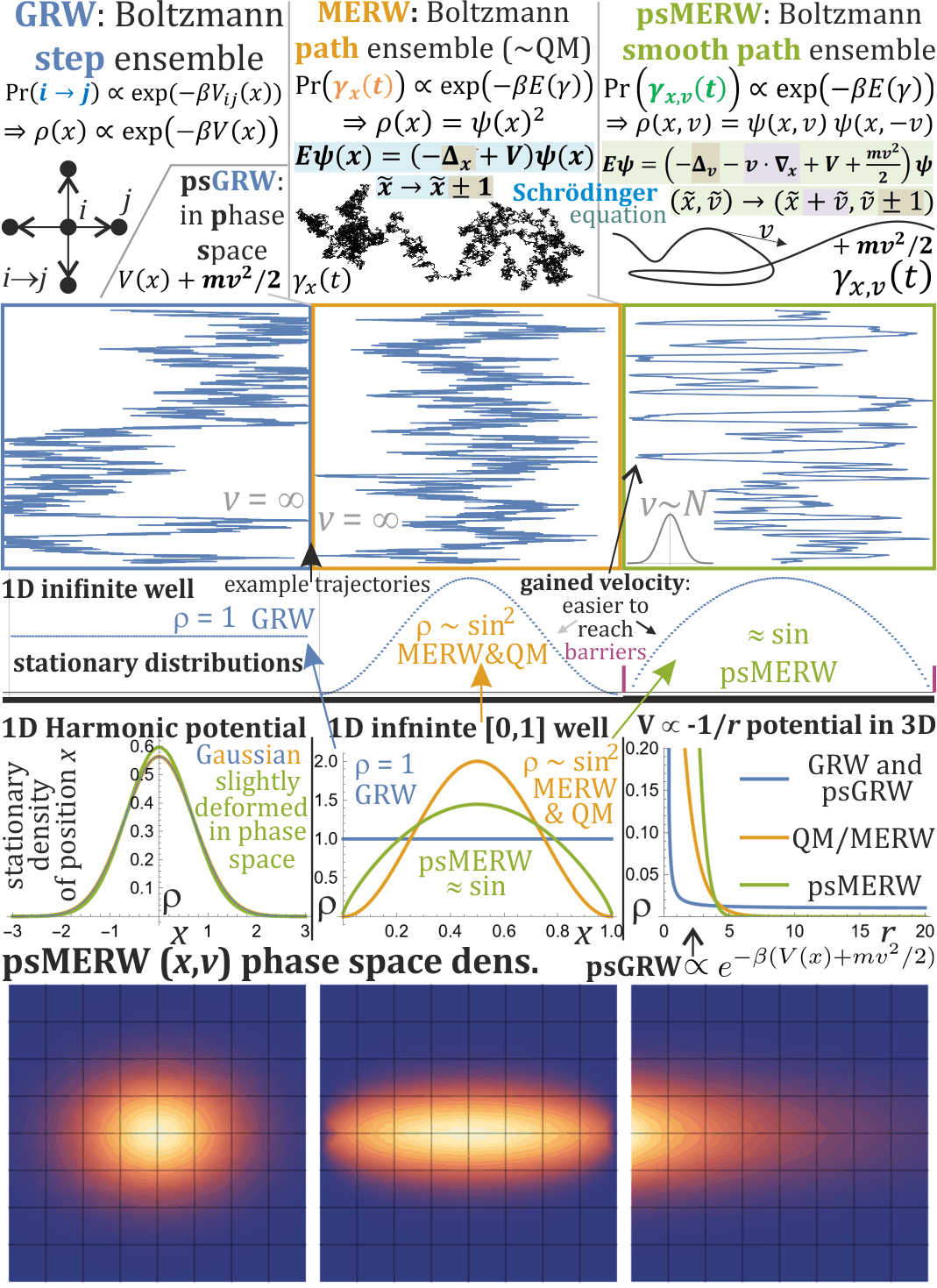}
        \caption{While (Jaynes) maximal entropy principle says we should use Boltzmann ensembles, being at heart of statistical physics, there are various ways to apply it, e.g. steps, paths and smooth paths ensembles. They lead to different predictions, this way bringing difficult question: which one is the most appropriate for various scenarios. The novel smooth path ensembles are obtained by going to phase space - with velocity controlling spatial step, and being randomly modified: $(\tilde{x},\tilde{v})\to (\tilde{x}+\tilde{v},\tilde{v}\pm 1)$. Both MERW and psMERW have stationary distributions being ground states of Schr{\"o}dinger equation, the latter using its written phase space version: with Laplacian $\Delta_v$ for velocity coming from $\tilde{v}\to\tilde{v}\pm 1$ transition. 
        There are shown example trajectories for 1D infinite potential well, generated with Mathematica code in \url{https://community.wolfram.com/groups/-/m/t/3124320}. We can see smooth paths are more likely to reach barriers thanks to randomly gained velocity $v$, which has distribution close to Gaussian.     
        In the bottom part there are shown stationary probability distributions: for all approaches in space (up) and for psMERW in phase space (down), for 3 popular potentials: 1D harmonic potential, 1D infinite well as above, and radial 3D $V\propto -1/r$ point electric or gravitational potential.} 
        \label{intr}
\end{figure}

\begin{figure}[t!]
    \centering
        \includegraphics{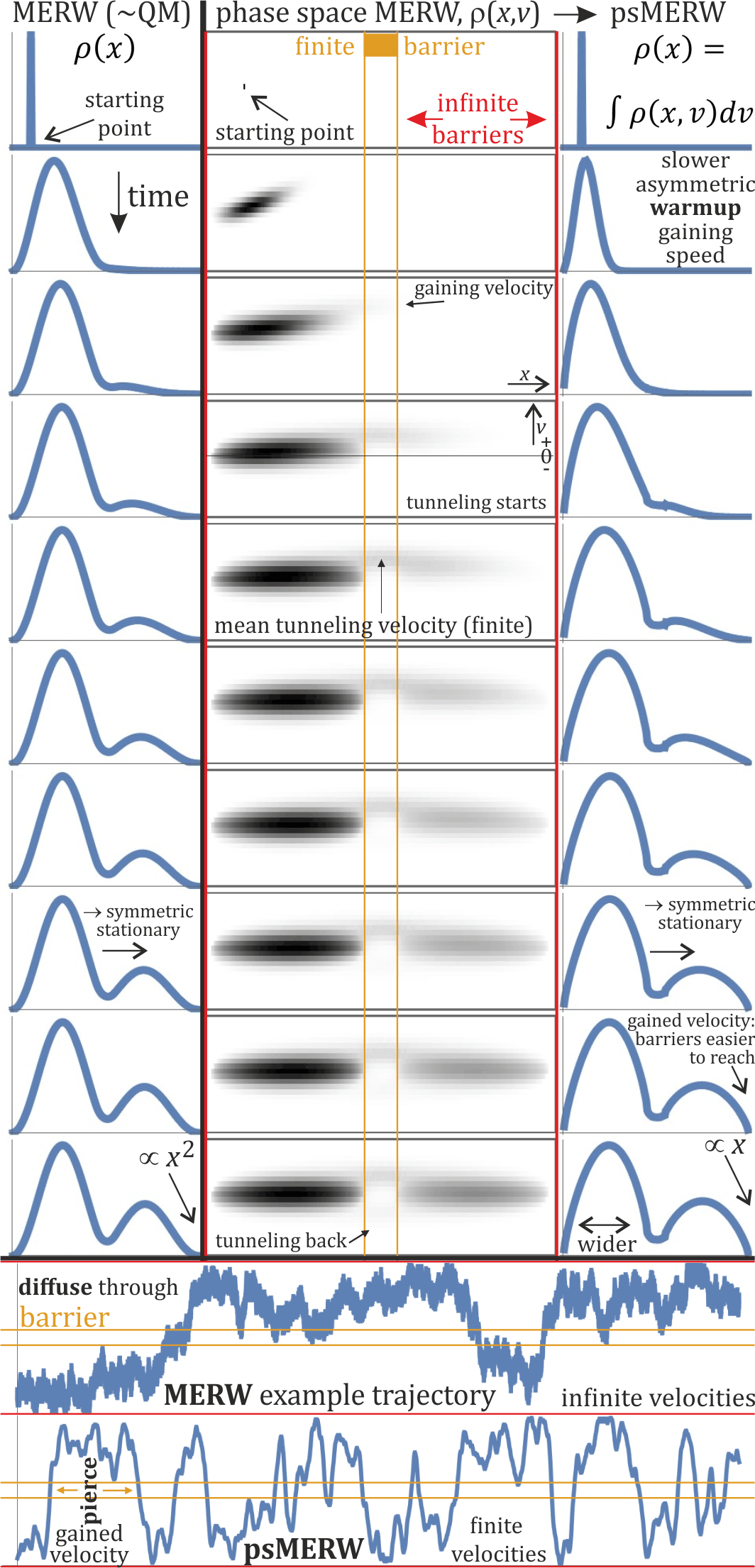}
        \caption{Tunneling comparison for dynamic scenario: example of discrete numerical evolution of probability distribution for MERW and psMERW on size 200 lattice in space (21 in velocity) and infinite potential well with additional symmetric potential barrier in the center. The parameters were chosen to get similar behavior - still obtaining visible qualitative differences, like boundary behavior (randomly gained velocity in psMERW makes it easier to reach barriers), differences in warmup, and widths of distributions. We can see velocity asymmetry during warmup - toward barrier have faster entropy growth. While MERW/QM tunneling velocity is a controversial question by some claimed to use superluminal~(\cite{tun, tun1}) - we see it rather diffuses through barrier, in phase space we get concrete finite velocities (like electrons in \cite{tun2}) - randomly gained velocity to pierce the barrier. Fig. \ref{tun} shows simulations of transition probability dependence from both barrier width, but also initial velocity. } 
        \label{tunnel}
\end{figure}

While such infinite velocity trajectories seem clearly nonphysical, a basic question is if physics really use them? Maybe only we use them to simplify mathematics? It is mathematically more difficult, but doable to use ensembles of more physical trajectories - bringing question if they could lead to a better agreement with experiment? The main purpose of this article is to start asking these basic but very difficult questions, by comparing 4 different approaches for a given spatial potential $V(x)$. Fig. \ref{intr} summarizes examples of predicted stationary probability distributions. For example uniform vs $\sim\sin^2$ vs $\sim\sin$-like behavior near a barrier from the infinite well case might be distinguishable. A dynamical example as tunneling with concrete final velocities is shown in Fig. \ref{tunnel}, and further in Section \ref{s5} - as expected getting exponential probability decrease with barrier width, but also dependence from the initial velocity, intuitively required but neglected in standard approaches. Basic code is available in the link on the right.

To consider ensembles of more physical trajectories, as in Langevin equation~\cite{langevin} we can go to $(x,v)$ phase space - include finite velocity $v$ in evolved state, which changes randomly and controls evolution in space $(x_{t+1}-x_t \propto v_t)$. To get various behaviors which could be compared experimentally, there is included general spatial potential $V(x)$.

Another basic crucial question is if we should base on maximizing local entropy production GRW (generic random walk) with walker considering single steps, what leads to Boltzmann ensemble in space. Or maybe better use maximizing mean entropy production MERW (maximal entropy random walk) using Boltzmann ensemble of entire paths - like random walk along Ising sequence, or "Wick-rotated" Feynman path ensembles, Euclidean quantum mechanics~\cite{zambrini} - mathematically leading to the same stationary probability distribution as quantum ground state, with much stronger Anderson-like localization property~\cite{MERWprl}.

This much stronger QM/MERW localization is crucial especially for electrons in semiconductor - standard diffusion would predict nearly uniform stationary electron probability distribution for such lattice of usually two types of atoms, which should flow if attaching external potential. In contrast, in QM/MERW and experiments these electrons are strongly localized, visualized e.g. in \cite{exp}, what prevents conductance in such semiconductor. MERW allows for working diffusions model of e.g. diode as semiconductor p-n junction~\cite{cond1}.

Neutrons are another objects confirmed to have quantum ground state stationary probability distribution~\cite{neutron} as predicted by both MERW and QM. Also for "walking droplets": classical objects with wave-particle duality, there were experimentally observed QM-like statistics~\cite{wavelike}. They brings a difficult general question which approach should be used in which case, e.g. for diffusion of molecules, solitons, or dust halos in astronomical settings.

While MERW might be appropriate for frequently interacting objects, e.g. for rarely interacting sparse dust halo it seems more connivent to consider MERW in phase space. Combining them was one of motivations for this article, also to compare predictions of all these models, which hopefully could be experimentally distinguished in some future.

Beside Langevin equation, related approaches are e.g. phase-space formulation of quantum mechanics~\cite{phasespace}, or Feynman path ensemble in phase space~\cite{fps}. However, they do not actually use velocities to choose spatial step, leading to different equations. While here we consider Boltzmann path ensembles in phase space, for Feynman it can be found in \cite{bouchaud} discussed further.

Unfortunately formulas for such more physical trajectories become more complicated, requiring to solve functional eigenequations, for which analytical solutions could be found rather only for very simple cases. Fortunately there are  available tools to do it numerically, like the used \verb"NDEigensystem" function of Wolfram Mathematica.

Section \ref{s2} introduces to GRW and MERW philosophies including continuous limit. Section \ref{s3} contains their proposed extension to phase space. This is early version of article, hopefully stimulating further research in this direction, especially toward experimental distinguishing of the discussed 4 approaches. Section \ref{s4} briefly discusses expansion to infinite variance steps: ensembles of L{\'e}vy flights in space and phase space. Section \ref{s5} discusses tunneling in this approach, especially dependence from initial velocity.

\section{Generic and maximal entropy random walk }\label{s2}
\begin{figure}[t!]
    \centering
        \includegraphics{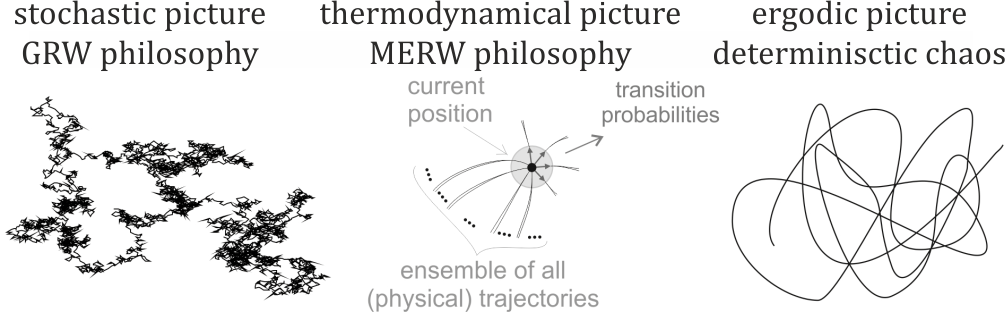}
        \caption{Three basic philosophies to predict stationary probability distributions. Left: in stochastic picture we basically guess stochastic propagator, usually maximizing entropy separately for each step (GRW), what leads to Boltzmann ensemble of positions in space. Right: in ergodic picture we assume we fully control deterministic evolution. Center: in contrast, in thermodynamics, statistical physics we do not assume knowing such details. Instead, we assume the safest statistical model accordingly to Jaynes' principle: maximizing entropy. It leads to: for each point consider Boltzmann ensemble of possible paths from it, their statics give MERW stochastic propagator. It is nonlocal (depends on the entire situation), but effective model - used only by us to predict the most likely distributions, not used by the object itself - governed by some complex unknown evolution. In this article we expand MERW philosophy to phase space, using Boltzmann ensemble of more physical trajectories. }
        \label{phil}
\end{figure}

\begin{figure}[b!]
    \centering
        \includegraphics{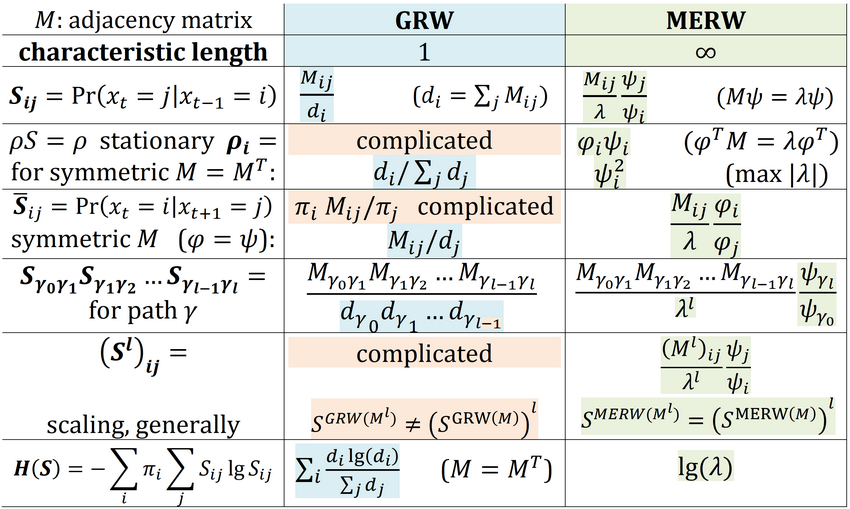}
        \caption{Gathered GRW, MERW formulas for graph defined by matrix $M$. }
        \label{table}
\end{figure}

This section briefly introduces to GRW/MERW\footnote{\scriptsize{MERW introduction: \url{https://community.wolfram.com/groups/-/m/t/2924355}}} random walks summarized in Fig. \ref{phil}, \ref{table}, for details see e.g. \cite{MERWprl, myphd}. 

While further we will use lattice for continuous limit, let us start with general random walk on a graph given by adjacency matrix $M$: $M_{ij}=0$ if there is no edge between $i$ and $j$, otherwise it is 1, or for generality it has some weight here. For Boltzmann path ensemble it is natural to parameterize: 
\be M_{ij}=\exp(-\beta V_{ij})\qquad (\textrm{e.g.}\ V_{ij}=(V_i+V_j)/2) \ee 
for $V_{ij}$ being energy of $i\to j$ step, and $\beta=1/k_B T$ in thermodynamics for $T$ temperature, for quantum-like statistics can be related with Planck's $\hbar$.

For such given $M$ matrix we would like to find stochastic matrix of transition probabilities for Markov process:
$$S_{ij}=\textrm{Pr}(X_t=j|X_{t-1}=i)\quad\textrm{satisfying}\quad \sum_j S_{ij}=1$$
and using only allowed transitions: $M_{ij}=0 \Rightarrow S_{ij}=0$, what is equivalent with assigning $V_{ij}=\infty$ energy to this transition.

A basic approach to choose $S$, referred as generic random walk (GRW), is assuming uniform/Boltzmann ensemble among single possible steps, this way maximizing local entropy for this step (minus mean energy for Boltzmann):
\be S^{GRM}_{ij}=\frac{M_{ij}}{d_i}\quad\textrm{for}\quad d_i=\sum_j M_{ij}\ee

For symmetric $M=M^T$ it leads to stationary probability distribution $\rho S=\rho$ being just $\rho_i^{GRW}= d_i/\sum_j d_j$. \\
\subsection{Maximal entropy random walk (MERW)}
In contrast, in MERW we maximize mean entropy production (minus mean energy for Boltzmann), what is equivalent to uniform/Boltzmann ensemble among infinite paths.

The basic tool to calculate its stationary distribution and propagator is noticing that for $M_{ij}=\exp(-\beta V_{ij})$, matrix power $M^l$ combinatorially contains sum of Boltzmann path ensemble:
\be (M^l)_{ij}=\sum_{\gamma:\gamma_0=i,\gamma_l=j}\exp(-\beta E(\gamma))\qquad \textrm{for}\ \ E(\gamma)=\sum_{i=1}^l V_{\gamma_{i-1}\gamma_i}\label{e2}\ee
We are interested in $l\to \infty$ limit, for which using natural assumption of connected and acyclic graph, the Frobenius-Perron theorem says there is a single dominant eigenvalue $\lambda$, allowing to use $\phi,\psi$  dominant eigenvectors in the limit:
\be M^l\propto \lambda^l \psi \phi^T \equiv \lambda^l |\psi\rangle \langle\phi|\qquad\textrm{for}\quad l\to \infty \label{e3}\ee 
$$ M\psi=\lambda \psi\qquad \phi^T M =\lambda \phi^T \quad\textrm{maximizing}\ |\lambda|$$

\begin{figure}[t!]
    \centering
        \includegraphics{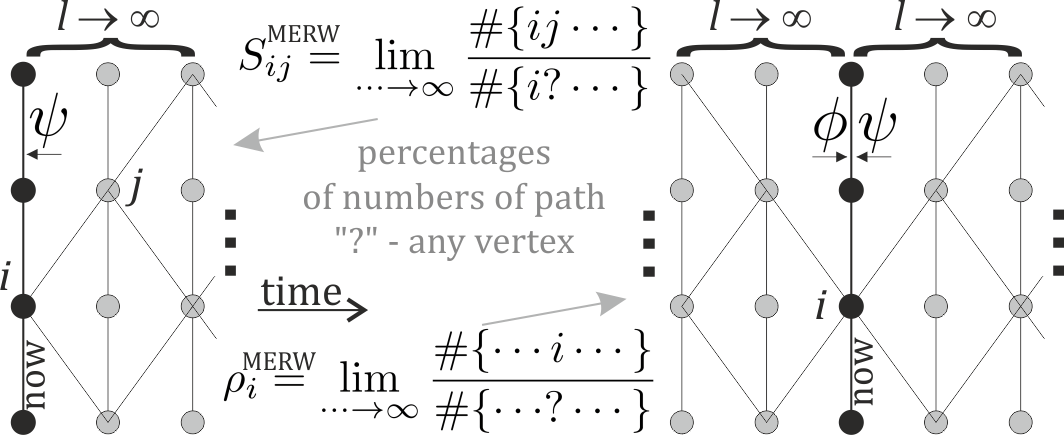}
        \caption{Diagrams for derivation of MERW formulas from path ensembles, exploiting that for $M_{ij}=\exp(-\beta V_{ij})$, matrix power $(M^l)_{ij}\approx \lambda^l \psi \phi^T$ combinatorially contains summation over uniform/Boltzmann ensemble of length $l$ paths from $i$ to $j$. To get stochastic propagator $S^{MERW}_{ij}$, we can consider length $l$ paths from vertex $i$, and look at distribution of their first step for $l\to \infty$ limit. To get stationary probability distribution we can multiply such two propagators: $\rho_i\propto\lim_{l\to\infty} (M^l)_{ji} (M^l)_{ik}\propto \phi_i \psi_i$.   }
        \label{MERWform}
\end{figure}

We can use both (\ref{e2}) and (\ref{e3}) to find propagator and stationary probability distribution (for any $j,k$) as in Fig. \ref{MERWform}:
\be S_{ij}^{MERW}=\frac{\textrm{Pr}(ij)}{\textrm{Pr}(i)}=\lim_{l\to\infty} \frac{M_{ij}(M^l)_{jk}}{(M^{l+1})_{ik}}=\frac{M_{ij}}{\lambda}\frac{\psi_j}{\psi_i}\ee
\be \rho^{MERW}_i \propto \lim_{\l\to\infty} (M^l)_{ji}(M^l)_{ik}\propto \phi_i\psi_i \label{Born} \ee
This way $\phi$ is distribution at the end of past half-paths, $\psi$ at the end of future half-paths. They come from $M^l$ or $(M^T)^l$ for $l\to \infty$ propagators from minus/plus infinity, differing by time direction. To get some position we need to randomly get it from both direction, so its probability is $\rho_i\propto \phi_i \psi_i$ product of two probabilities. For (time) symmetric $M=M^T$ both eigenvectors are equal: $\phi=\psi$, getting $\rho_i \propto \psi_i^2$ Born rule.

\begin{figure*}[h!]
    \centering
        \includegraphics[width=17cm]{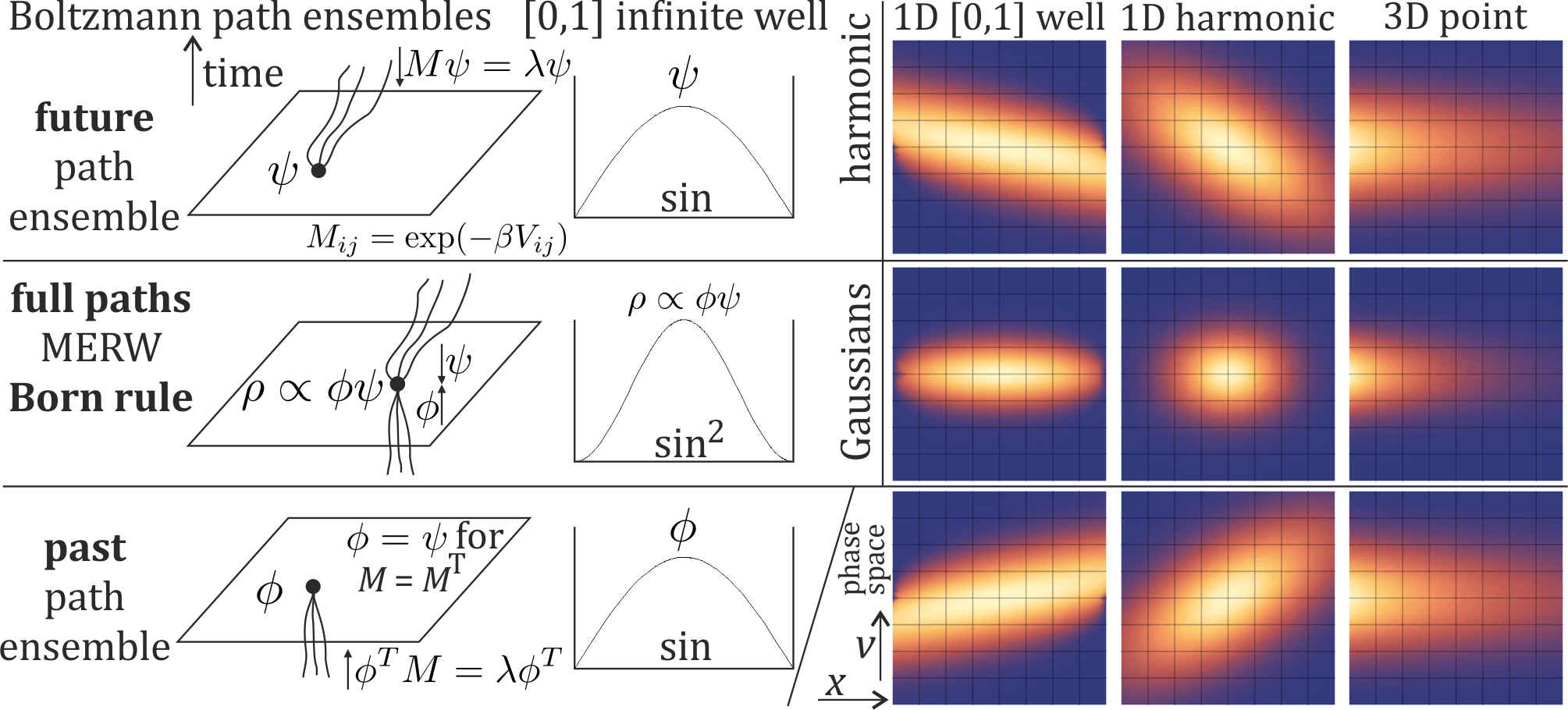}
        \caption{Boltzmann path ensembles: $\psi$ is probability distribution at the end of future half-paths, $\phi$ at the end of past half-paths. They can be viewed as results of propagator from plus/minus infinity: $M^l$, $(M^T)^l$ for $l\to \infty$. To get stationary probability distribution in the center of full paths we need to randomly get it from both directions: $\rho=\phi \psi$ Born rule. While for (time) symmetric $M=M^T$ both are equal $\phi=\psi$, in the discussed phase space case, time symmetry inverses velocity - on the right there are 3 cases as in Fig. \ref{intr}, now with shown $\psi(x,v)=\phi(x,-v)$, which differ by inversion of velocity, making $\rho=\phi \psi$ symmetric. This symmetry is required as inverting all paths in their Feynman ensemble (including velocities), we get the same ensemble. While here for Boltzmann path ensembles we need two real eigenfunctions, in Feynman path ensembles they are combined in real/imaginary parts of single complex eigenfunction~\cite{bouchaud}.
         }
        \label{paths}
\end{figure*}

\subsection{Lattice and MERW continuous limit to Schr{\"o}dinger equation}
Let us discretize space $\mathbb{R}^D\ni x=\delta \xb$ for $\xb\in\mathbb{Z}^D$ and time $\mathbb{R}\ni t=\epsilon \tb$ for $\tb\in \mathbb{Z}$. For continuous limit $\epsilon,\delta\to 0$. Also discretize spatial potential: $V_{\xb}\equiv V(\delta \xb)$. For simplicity let us work in $D=1$, with mentioned generalization for $D>1$.

\subsubsection{GRW leading to Boltzmann space ensemble}
 Allowing only jumps up to the nearest neighbors: step $s\in \{-1,0,1\}$, we can choose $M$ matrix as symmetric and using potential:
\be M_{\xb,\xb+s} = \exp(-\beta(V_{\xb}+V_{\xb+s})/2)\quad\textrm{for }s\in\{-1,0,1\}\label{Mdef}\ee
as it is symmetric, GRW stationary probability distribution is $\rho_{\xb}\propto M_{\xb,\xb- 1}+M_{\xb,\xb}+ M_{\xb,\xb+ 1}$. Assuming continuous potential $V(x)$, in the limit we get just Boltzmann ensemble in space:
\be \rho^{GRW}(x) \propto \exp(-\beta V(x)) \ee

\subsubsection{MERW leading to Schr{\"o}dinger ground state} For MERW we assume Boltzmann ensemble among paths, with path energy as sum of energies of single transitions. For continuous limit this energy should become integral over time $E(\gamma)=\int V(\gamma(t)) dt$, what requires to include time step $\epsilon$ in transition energy:
$$ M_{\xb,\xb+s} = \exp(-\beta\epsilon((V_{\xb}+V_{\xb+s})/2)$$
For MERW we first need to find the dominant eigenvector of adjacency matrix $\lambda\psi=M\psi$ for $M$ given by (\ref{Mdef}):
\be \lambda \psi_{\xb}=(M\psi)_{\xb}=\sum_{s\in\{-1,0,1\}}e^{-\beta\epsilon(V_{\xb}+V_{\xb+s})/2}\ \psi_{\xb+s}\ee
Taking $\exp(-\epsilon)\approx 1-\epsilon$ first order expansion and assuming continuous potential $V(x)$, for continuous limit we can approximate the above with:
$$ \lambda \psi_{\xb}\approx \psi_{\xb-1}+\psi_{\xb}+\psi_{\xb+1}-3\beta\epsilon V_{\xb}
$$
Subtracting $3\psi_{\xb}$ from both sides, and multiplying by $-1/3\beta\epsilon$:
$$\frac{3-\lambda}{3\beta\epsilon}\psi_{\xb} \approx -\frac{1}{3\beta} \frac{\psi_{\xb-1}-2\psi_{\xb}+\psi_{\xb+1}}{\epsilon} + V_{\xb} \psi_{\xb}$$
Due to change of sign, maximization of eigenvalue $|\lambda|$ becomes minimization of $(3-\lambda)/(3\beta\epsilon)$, which in continuous limit should converge to some energy $E$. Assuming $\epsilon =\delta^2$, the difference term will lead to Laplacian, hence in continuous limit we can write the eigenequation as stationary Schr{\"o}dinger equitation searching for the lowest energy $E$ ground state: 
\be E\psi =-C\Delta \psi +V\psi\qquad\quad \rho^{MERW}(x)\propto (\psi(x))^2\label{schr}\ee
for $C=1/3\beta$ in 1D or generally $1/(2D+1)$ in dimension $D$. We can choose parameters to make it $C=\hbar^2/2m$ as in standard Schr{\"o}dinger equitation, what is explored e.g. in ~\cite{myphd} or in \cite{darwin} finding also some agreement toward Dirac equation.
\section{Random walks in phase space} \label{s3}
While standard random walks or Feynman path ensembles use nowhere differentiable paths of infinite kinetic energy, here we would like use more physical paths by going to phase space $(x,v)$: with random change of (finite) velocity $v$, which defines deterministic change of position $x$.

Like in popular Langevin approach~\cite{langevin}:
$$ m\frac{dx}{dt}=v\qquad\qquad m\frac{dv}{dt}=-\Gamma v +\eta $$
for $\Gamma$ describing damping (neglected in current version). It has no potential, hence independent finite variance infinitesimal steps would lead to Gaussian distribution for velocities.

Here we would like to include energy - both kinetic $mv^2/2$ and some general position dependant $V(x)$:
\be \bar{V}(x,v)=V(x)+ m\|v\|^2/2\ee
\subsection{GRW phase space continuous limit (psGRW)}
Discretizing  $\mathbb{Z}\ni \xb =x/\delta$ position, and  $\mathbb{Z}\ni \vb =v/\zeta$ velocity, GRW as Boltzmann ensemble of single steps e.g. $s,s'\in \{-1,0,1\}$ can be chosen as:
$$ S_{(\xb,\vb),(\xb+\vb,\vb+s)}=\frac{\exp(-\beta(V(\delta \xb)+\frac{1}{2}m (\zeta (\vb+s))^2))}{\sum_{s'} \exp(-\beta(V(\delta \xb)+\frac{1}{2}m (\zeta (\vb+s'))^2))} $$
The spatial potential contributions $V(\delta \xb)$ cancel out - leaving random walk of velocity alone, in $mv^2/2$ kinetic energy acting as harmonic potential for velocity. 

Therefore, while $M$ is not symmetric due to velocity inversion in $M^T$, for velocity alone it can be made symmetric, leading to $\rho(v)\propto \exp(-\beta mv^2/2)$ Gaussian stationary probability distribution for velocity, independent from position.

For such position independent symmetric Gaussian velocity distributions, in infinitesimal limit we can get symmetric $M$ among positions, allowing to conclude psGRW stationary probability distributions as Boltzmann distribution in the phase space:
\be \rho^{psGRW}(x,v) \propto \exp(-\beta (V(x)+mv^2/2))\rho\ee
\subsection{MERW phase space continuous limit (psMERW) }
\subsubsection{Derivation} Let us discretize space $\mathbb{R}^D\ni x=\delta \xb$ for $\xb\in\mathbb{Z}^D$ and time $\mathbb{R}\ni t=\epsilon \tb$ for $\tb\in \mathbb{Z}$. For continuous limit we will take $\epsilon,\delta\to 0$. Discretized $\vb\in\mathbb{Z}^D$ velocity for $\xb\to \xb+\vb$ transition in single step corresponds to $v=\frac{\delta}{\epsilon}\tilde{v}\in\mathbb{R}^D$ real velocity. As previously we calculate for $D=1$, then mention for general $D$.

Denote $\psi_{\xb,\vb}=\psi(\delta \xb, \frac{\delta}{\epsilon} \vb)=\psi(x,v)$ as discretization of continuous eigenfunction in phase space. Analogously for potential: $\bar{V}_{\xb,\vb}=\bar{V}(\delta \xb, \frac{\delta}{\epsilon} \vb)=\bar{V}(x,v)$.

In one step $\tilde{x}\to\xb+\vb$, and there is a random velocity change $\vb\to \vb+s$, we can assume $s\in\{-1,0,1\}$. To include damping in future, we could add velocity reduction in below step. In 1D the MERW eigenequation after neglecting $O(\epsilon^2)$ terms becomes:
\be\lambda \psi_{\xb,\vb}=(M\psi)_{\xb,\vb}=\sum_{s\in\{-1,0,1\}} e^{-\epsilon \beta \bar{V}_{\xb+\vb,\vb+s}}\psi_{\xb+\vb,\vb+s}\approx \label{eigp}\ee
$$\approx \sum_{s\in\{-1,0,1\}} (1-\epsilon \beta \bar{V}_{\xb,\vb})(\psi_{\xb,\vb+s}+\epsilon v (\nabla_x\psi)_{\xb,\vb}) $$
thanks to approximation with derivative $\nabla_x\equiv \partial/\partial x$: $$\psi_{\xb+\vb,\vb}=\psi(\delta (\xb+\vb),v)=\psi(x+\epsilon v,v)\approx \psi_{\xb,\vb} +\epsilon v(\nabla_x\psi)_{\xb,\vb}$$
Neglecting $O(\epsilon^2)$ terms eigenequation (\ref{eigp}) becomes:
$$\lambda \psi_{\xb,\vb}\approx
\psi_{\xb,\vb-1}+\psi_{\xb,\vb}+\psi_{\xb,\vb+1}+
3\epsilon v(\nabla_x\psi)_{\xb,\vb}
-3\beta \epsilon \bar{V}_{\xb,\vb} \psi_{\xb,\vb}$$
Now subtract $3 \psi_{\xb,\vb}$ from both sides and multiply by $-1/(3\beta (\delta/\epsilon)^2)$, getting:
$$\frac{3-\lambda}{3 \beta(\delta/\epsilon)^2}\psi_{\xb,\vb}\approx 
-\frac{1}{3\beta}\frac{\psi_{\xb,\vb-1}-2\psi_{\xb,\vb}+\psi_{\xb,\vb+1}}{(\delta/\epsilon)^2}-$$
$$-v\frac{1}{\beta}\frac{\epsilon}{(\delta/\epsilon)^2} (\nabla_x\psi)_{\xb,\vb}+\frac{\epsilon}{(\delta/\epsilon)^2} \bar{V}_{\xb,\vb} \psi_{\xb,\vb}$$
Assuming $\delta/\epsilon\to 0$, the discrete Laplacian tends to continuous $\Delta_v$ for velocity. As previously we would like to interpret the first coefficient as energy: $\frac{3-\lambda}{3 \beta(\delta/\epsilon)^2}\to E $.

For $\nabla_x$ gradient term we can choose: 
\be\frac{1}{\beta} \frac{\epsilon}{(\delta/\epsilon)^2}\to \xi \qquad \textrm{for example}\quad \epsilon=\sqrt[3]{\beta \xi\, \delta^2}\ee
or use a sequence $\xi \to 0$ for $\xi=0$ in continuous case.

Finally in the continuous limit we get \textbf{stationary phase space Schr{\"o}dinger equation} for  $\psi\equiv\psi(x,v)$:
\be E\psi=-C\Delta_v \psi- \xi v\cdot \nabla_x \psi + \left(V+\frac{1}{2} m\|v\|^2\right)\psi \label{fin}\ee
where $C=\frac{1}{3\beta}$, or generally $C=\frac{1}{(2D+1)\beta}$ in dimension $D$, is some constant which can be directly chosen, e.g. dependent on $\hbar$ or temperature. Analogously for $\xi$, down to $\xi\to 0$ limit making position and velocity independent.

To find the stationary probability distribution $\rho(x,v)$, we need to solve eigenequation (\ref{fin}) minimizing energy $E$, what corresponds to $|\lambda|$ maximization. Higher energy contributions have exponential decay here. This time $M$ is not symmetric as $M^T$ time symmetry inverses velocity, so we should find left/right eigenfunction. Fortunately, as in Fig. \ref{paths}, they differ by just change of sign of velocity $\phi(x,v)=\psi(x,-v)$, getting Born rule:
\be \rho(x,v)\propto \psi(x,v)\ \psi(x,-v)\ee
\subsubsection{Comparison with Feynman path ensemble}
Similar $t\to it$ "Wick rotated" phase space Schr{\"o}dinger equation was earlier derived by Bouchaud~\cite{bouchaud} for QM, Feynman path ensembles:
$$i\hbar \frac{\partial \Psi(x,q)}{\partial t}=\frac{\hbar}{2\tau_c}
[\Delta_q \Psi(x,q) -q^2 \Psi(x,q)]+$$
\be\qquad +i \sqrt{\frac{\hbar^3}{m\tau_c}} q\nabla_x \Psi(x,q)+
V(x)\Psi(x,q)\label{pss}\ee
for $\tau_c$ momentum correlation time: as in Ornstein-Uhlenbeck process assuming $\langle p(s)p(u)\rangle=\frac{m\hbar}{2\tau_c}\exp(-|s-u|/\tau_c)$, and rescaled momentum $p=\sqrt{m\hbar/\tau_c }q$. Comparing it with (\ref{fin}) suggests parameter choice for quantum scale applications. 

Beside $t\to it$ "Wick rotation to imaginary time", there are some subtle differences between Boltzmann and Feynman path ensembles. Hamiltonian corresponds to minus $M$ matrix, energy $E$ corresponds to minus $\lambda$ - discussed here maximization of $\lambda$ corresponds to minimization of energy for the ground state. 

In Boltzmann path ensemble excited states vanish exponentially: $M\psi^k =\lambda_k \psi^k$ for $|\lambda_0|\geq |\lambda_1|\geq \ldots$, then after $t$ steps $\sum_k a_k \psi^k \to \sum_k (\lambda_k/\lambda_0)^t a_k \psi^k$ (+normalization). In contrast, in QM/Feynman path ensembles $\psi^k$ excited states are stable, rotate in complex plane. Excited e.g. atoms are believed to deexcite through interaction with environment.

Transposition $M^T$ corresponds to time reversal. As in Fig. \ref{paths}, for non-symmetric $M$ ensembles of future and past half-paths lead to different $\phi,\psi$ which are real for Boltzmann ensembles. In contrast, (\ref{pss}) Hamiltonian is self-adjoint: invariant under time symmetry in conjugation, containing such two asymmetric real eigenfunction in real/imaginary parts of one complex wavefunction.\\

\subsubsection{Calculation remarks}
While for QM/MERW there are well known analytical solutions of such stationary Schr{\"o}dinger equation e.g. for harmonic potential, infinite well, point $V\propto -1/r$ potential, their phase space version are more complicated and Wolfram Mathematica \verb"DEigensystem" was not able to find analytical solutions, hence for their example solution in Fig. \ref{intr} there was used numerical \verb"NDEigensystem", also required for more complex settings. It uses finite element method, and already had various issues in the discussed basic cases - resolved e.g. by manually specifying Dirchlet conditions for infinite well, or for point potential: solving for $\psi(x,v)\sim \exp(-v^2-x) \phi(x,v)$ type substitution assuming asymptotic behavior. 

Alternative approaches are discretization like in Fig. \ref{tunnel}, or approximation by solving in a finite orthogonal basis of functions (removing higher modes), especially Gaussian distribution times Hermite polynomial for velocities (ground state already gets such higher modes), further developed in \cite{bouchaud}. Numerical methods for psMERW will require further work, especially to include interaction between multiple particles.

\section{Boltzmann ensembles of L{\'e}vy flights} \label{s4}
Laplacians in the discussed derivations came from infinitesimal behavior of $x\to x\pm 1$ (MERW) or $v\to v\pm 1$ (psMERW) random steps. As these random steps have finite variance, the central limit theorem says they lead to Gaussian distribution, which can be deformed here through Boltzmann path ensemble.

Without the assumption of finite variance, there is generalized central limit theorem~\cite{levy1939}, which in infinitesimal limit replaces Laplacian with its Riesz fractional derivative generalization, leading to L{\'e}vy flights~\cite{mandelbrot} with steps of infinite variance.

Specifically, $\varphi_X(k)\equiv F_k[X]=E[e^{ikX}]=\int_{-\infty}^\infty e^{ikx} dF_X(x)$ characteristic function for summation of independent random variables is just product: $\varphi_{X+Y}(k)=E[e^{ikX}e^{ikY}]=\varphi_{X}(k)+\varphi_{Y}(k)$. Hence adding multiple independent steps from the same distribution (i.i.d.) results in taking powers of characteristic functions, in the limit of infinite number of steps leading to below characteristic function of stable distribution centered in $\delta$:
\be \varphi_{\alpha s}(k)=\exp\left(it\delta -\|s k\|^\alpha \right) \ee
generally also with asymmetry term, usually neglected in L{\'e}vy flights (maybe it should be considered?). For $\alpha=2$ it is just Gaussian distribution, for $\alpha=1$ it is Cauchy distribution. For $\alpha<2$ it has infinite variance, and $\rho(x)\sim |x|^{-1-\alpha}$ tails. 

While $x\to x\pm 1$ has lead to Laplacian in infinitesimal limit, in L{\'e}vy flights we replace such finite variance term with of $\exp\left(-\|k\|^\alpha \right)$ characteristic function - leading to fractional Riesz derivative, Laplacian, in practice defined for characteristic function:
\be F_k \left[(-\Delta)^{\alpha/2} \varphi \right] =\|k\|^\alpha F_k[\varphi]\ee
where $\|\cdot\|\equiv \|\cdot\|_2$ is Euclidean norm - it is crucial to use spherically symmetric. Using coordinate-wise $|k|^\alpha$ instead, would lead to dominant steps in canonical directions.

For MERW as Boltzmann path ensemble, to use ensemble of L{\'e}vy trajectories of infinite step variance, we need to replace Laplacian in Schr{\"o}dinger equation wits such fractional one $(-\Delta_x)^{\alpha/2}$.  Such \textbf{fractional Schr{\"o}dinger equation} is considered in literature~\cite{fsch}, also recently getting experimental realization~\cite{fsch1}. For example in 1D infinite potential well, in MERW we get $E F_k[\psi]=|k|^\alpha F_k[\psi]$ type eigenequation, due to Dirchlet conditions having $\sin$-like solutions - going to Boltzmann ensemble of Levy trajectories does not change stationary probability distribution, only energy levels.

For psMERW analogously we can use L{\'e}vy flights of velocities by replacing Laplacian in (\ref{fin}) with fractional $(-\Delta_v)^{\alpha/2}$, getting \textbf{phase space fractional Schr{\"o}dinger equation}. It would allow to model ensembles of smooth paths with changes of velocity having infinite variance, for example corresponding to large velocity changes during collisions in e.g. dust. Applying Fourier transform to velocity $\tilde{\psi}\equiv \tilde{\psi}(x,k)=\int e^{ikv} \psi(x,v)$ it becomes: 
$$ E\tilde{\psi}=(C|k|^\alpha+V) \tilde{\psi} + \int e^{ikv}\left( \frac{m}{2}v^2 \psi - \xi v\partial_x \psi\right) dv$$
\be E\tilde{\psi}=(C|k|^\alpha+V) \tilde{\psi} - \frac{m}{2}\partial_{kk} \tilde{\psi} + i\xi \partial_{xk} \tilde{\psi} \ee

\begin{figure}[t!]
    \centering
        \includegraphics{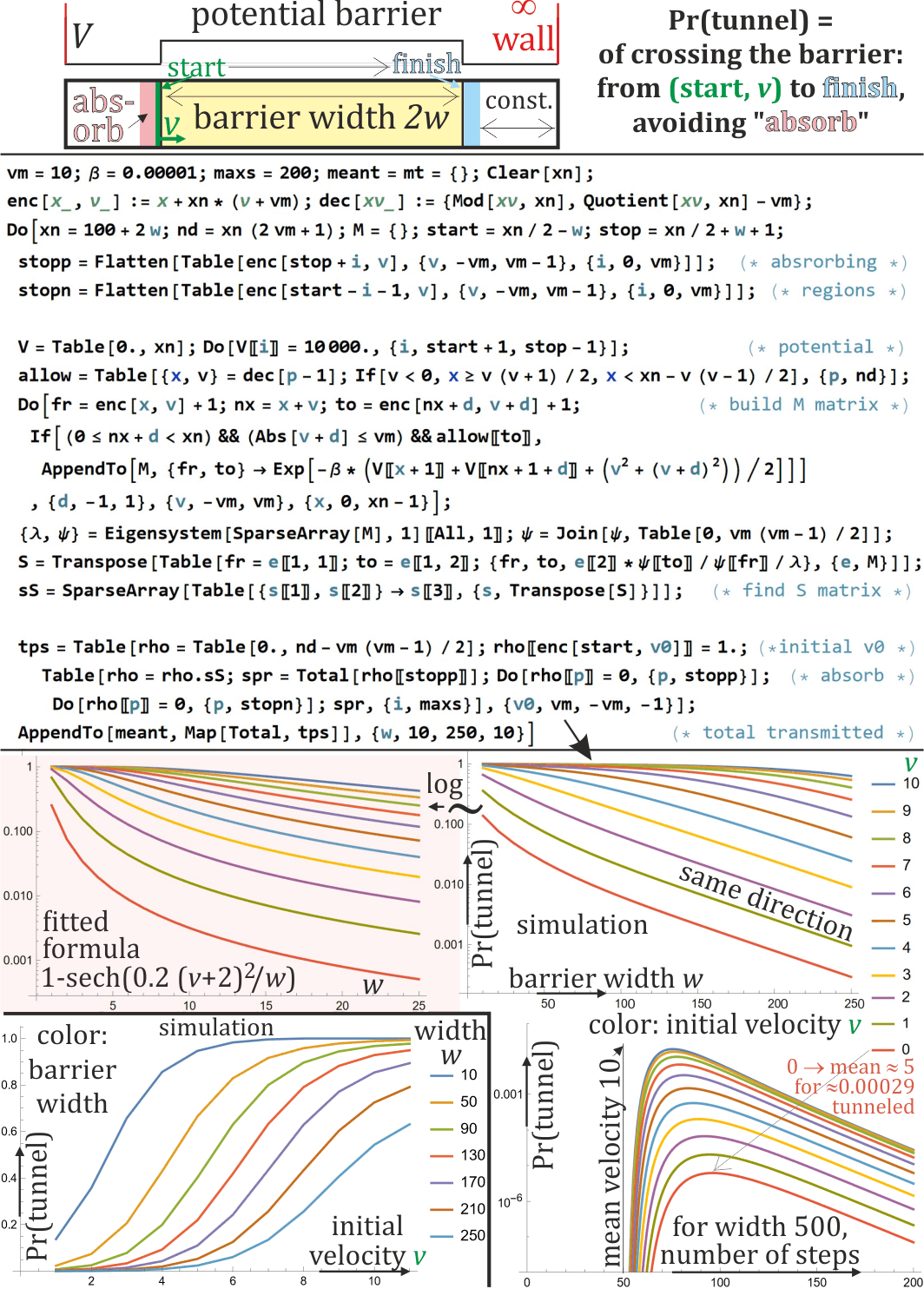}
        \caption{Tunneling investigation setting (top), Mathematica source (center), and simulation results (bottom) of tunneling probability dependence from barrier width $w$ and initial velocity $v$ (and number of steps in bottom-right). The (discrete) setting is potential barrier of varying width, with symmetric two zero potential regions of the same fixed width on both sides (ending with infinite potential walls). We start with probability distribution concentrated in a single point just before the barrier (green start), and chosen initial velocity (working in phase space). Then we evolve this distributions using the found psMERW stochastic propagator, zeroing all density which visited the two absorption regions (absorb and finish), earlier storing probabilities which ended in the finish region. There are shown plots for fixed initial velocity (upper-right) and on its left using rough approximation with guessed $\sim 1-\textrm{sech}(v^2/w)$ formula. In logarithmic scale we can observe constant asymptotic direction: exponent of transition probability reduction with barrier width $w$. Bottom-left plot fixes barrier width instead, showing transition probability approaching 1 for high initial velocities. Similar plots were found in 2009 in walking droplet tunneling experiment \cite{eddi2009,nachbin2017,tadrist2020}. Bottom-right plot fixes barrier width $2w=500$, showing probability of reaching the finish in a given number of steps - e.g. allowing to estimate mean velocity: lowered for fast initial velocity, increased for low initial.  }
        \label{tun}
\end{figure}

\section{Tunneling simulations with initial velocity} \label{s5}
As in \ref{tunnel}, we can simulate tunneling with the discussed path ensembles - for example using infinite potential well with additional barrier inside, e.g. rectangular assumed here. 

Additionally, for massive particles with high initial velocity, intuitively probability of crossing the barrier should approach 1. However, in standard Feynman/Boltzmann path ensembles (QM/MERW) this velocity is infinite, we cannot control it. In contrast, with discussed phase space ensembles we can choose initial velocity - Fig. \ref{tun} shows code and simulation results, asymptotically getting exponential probability decrease with barrier width, but also transition probability approaching 1 for high initial velocities. We can also see mean velocity decreased for high initial (e.g. $10 \to \approx 7$)  and increased for low (e.g. $0 \to \approx 5$) - the latter for a tiny fraction ($\approx 0.00029$) of walkers being successful, the remaining were absorbed instead.

In literature, looking similar dependencies were found especially in 2009 tunneling experiment for walking droplets~\cite{eddi2009,nachbin2017,tadrist2020}. They complement classical behavior with history dependence through field carrying (pilot-like) waves from the previous bounces - high complexity of this dependence leads to practically random transition probabilities. Not knowing details of such hidden field, from statistical physics perspective it seems natural to assume effective description by Boltzmann path ensembles - requiring smooth paths: phase space variants.

This behavior can be approximated by ordinary differential equations - considered for tunneling in \cite{Alvaro}. However, it leads to much more complicated behavior, currently not getting the required exponential decrease of transition probabilities with barrier width.

The details of such initial velocity dependence of tunneling probability were not found in literature - requiring further investigation, maybe deriving from the discussed smooth path ensembles, phase space Schr{\"o}dinger equation, or trying to fit the numerical simulations - which is close to $\approx 1-\textrm{sech}(a v^2/w)$ type dependence for some $a$ parameter. 

\section{Conclusions and further work}
While popularly considered path ensembles usually use infinite kinetic energy paths, it is mathematically more difficult but doable to use ensembles of more physical paths instead by going to phase space, however, it leads to slightly different predictions. This article discusses some basics with included spatial potential to bring attention, hopefully leading to some attempts to experimentally determine the most appropriate ones for various physical settings.

This is early article opening various directions for further work, e.g.:
\begin{itemize}
  \item Search for possibilities of experimental determination which of the discussed approaches is the most appropriate for various scenarios, especially using stationary probability distributions of positions, maybe also velocities e.g. from redshifts, or in dynamical settings like tunneling - in various situations from microscopic e.g. neutrons~\cite{neutron}, molecules, solitons~\cite{wavelike}, to astronomical e.g. dust halos.
  \item Find parameters, understand their dependence, universality.
  \item Mathematical improvements, especially of numerical approaches, search for analytical formulas for basic cases, maybe include damping like in Langevin equation, larger velocity changes through collisions, relativistic corrections, magnetic field through vector potential, etc.
  \item While for simplicity we have discussed only stationary situation, it is worth to also consider dynamics - both evolution in fixed potential like for electron conductance~\cite{cond1} or tunneling in Fig. \ref{tunnel}, but also much more difficult case of varying potential - with basics discussed in \cite{myphd}.
  \item While the discussed analysis was for a single walker, in practice we usually have multiple interacting - what should be finally included. One way is through mean-field treatment assuming the remaining have the same distribution, used e.g. for MERW electron conductance model~\cite{cond1} to include potential contributions from the remaining electrons. For psMERW it could be useful e.g. for astronomical dust halos to include their gravitational self-interaction.  
  \item While we have focused on ensembles of full paths, maybe it is also valuable to consider ensembles of unidirectional paths, which have asymmetric velocity distributions like in Fig. \ref{paths}. 
  \item While we have focused on continuous limit of infinitesimal lattice constants, finite lattice can be considered e.g. for conductance models imagining electrons jumping between atoms in a lattice. Also, for MERW Darwin term was recently derived~\cite{darwin} as correction from use of finite lattice. Especially the latter suggests to also consider psMERW in some finite lattice, and closely look at corrections it brings.  
  \item Investigate tunneling probability dependence - not only from barrier widths, but also initial velocity. 
\end{itemize}

\bibliographystyle{IEEEtran}
\bibliography{cites}
\end{document}